\documentclass[preprint,5p,times,twocolumn,12pt]{elsarticle}

\usepackage{ifpdf}
\usepackage{graphicx,natbib,lineno,color}
\ifpdf
\usepackage{amssymb}
\usepackage{lipsum}
\usepackage{bm}
\usepackage{amssymb}
\usepackage{pifont}
\usepackage{slashed}

\usepackage[%
breaklinks=true,%
colorlinks=true,%
linkcolor=blue,anchorcolor=blue,%
citecolor=blue,filecolor=blue,%
menucolor=blue,pagecolor=blue,%
urlcolor=blue]{hyperref}

\makeatletter
\def\elsartstyle{%
	\def\normalsize{\@setfontsize\normalsize\@xiipt{14.5}}
	\def\small{\@setfontsize\small\@xipt{13.6}}
	\let\footnotesize=\small
	\def\large{\@setfontsize\large\@xivpt{18}}
	\def\Large{\@setfontsize\Large\@xviipt{22}}
	\skip\@mpfootins = 18\p@ \@plus 2\p@
	\normalsize
}
\@ifundefined{square}{}{\let\Box\square}
\makeatother

\journal{Physics Letters B}
\def\date{ } 

\begin{document}
	
\begin{frontmatter}
	
\title{{\bf Rastall gravity is equivalent to Einstein gravity}}

\author{Matt Visser}

\address{School of Mathematics and Statistics,
Victoria University of Wellington;
PO Box 600, Wellington 6140, New Zealand.}

\ead{matt.visser@sms.vuw.ac.nz\qquad} 

\begin{abstract}
Rastall gravity, originally developed in 1972,  is currently undergoing a significant surge in popularity.
Rastall gravity purports to be a modified theory of gravity, with a non-conserved stress-energy tensor, and an unusual non-minimal coupling between matter and geometry, the Rastall stress-energy satisfying $[T_\mathrm{\,\sc R}]^{ab}{}_{;b} = {\lambda\over 4} \,g^{ab}\, R_{;b}$. 
Unfortunately, a deeper look shows that Rastall gravity is completely equivalent to Einstein gravity --- usual general relativity. The gravity sector is completely standard, based as usual on the Einstein tensor, while in the matter sector Rastall's stress-energy tensor corresponds to an artificially isolated \emph{part} of the physical conserved stress-energy.

\medskip\noindent
\emph{Date:} 30 November 2017; 8 December 2017; 17 April 2018; 9 May 2018; LaTeX-ed \today.

\medskip\noindent
\emph{Preprinted as:}  arXiv: 1711.11500 [gr-qc]
%
\end{abstract}


\end{frontmatter}
\parindent0pt
\parskip7pt
\section{Introduction}\label{S:intro}

Rastall gravity~\cite{Rastall:1972}, despite its somewhat mixed 45-year history, is currently undergoing a significant surge in popularity. Some 19 closely related articles have appeared so far in 2017~\cite{MoraisGraca:2017,Kumar:2017,Fabris:2017,Xu:2017,Darabi:2017,Moradpour:2017,Lobo:2017,Spallucci:2017,Koorambas:2017,Haghani:2017,Ma:2017,Licata:2017,Moradpour:2017b,Moradpour:2017c,Heydarzade:2017,Shabani:2017,Shabani:2017b,Santos:2017,Bronnikov:2017}. (See also~\cite{Fabris:1998,Fabris:2011,Batista:2011}.)

Unfortunately, as I shall argue below, Rastall gravity is completely equivalent to standard Einstein gravity --- general relativity --- all that is going on is that one is artificially splitting the physical conserved stress-energy tensor into two non-conserved pieces.

Historically, in 1972 Rastall tentatively suggested~\cite{Rastall:1972} that it might prove profitable to consider a covariantly non-conserved stress-energy tensor, one with
$\nabla_b [T_\mathrm{\,\sc R}]^{ab} \neq 0$.
In particular, he then suggested the phenomenological model $\nabla_b [T_\mathrm{\,\sc R}]^{ab} = F^a$, where $F^a$ is some vector field vanishing in flat spacetime. 

A (somewhat weak) plausibility argument then led him to consider
$\nabla_b [T_\mathrm{\,\sc R}]^{ab} \propto g^{ab} \nabla_a R$. 
Ultimately Rastall posited the existence of a constant $\lambda$ such that
for Rastall's non-conserved stress energy tensor
\begin{equation}
\nabla_b [T_\mathrm{\,\sc R}]^{ab} = {\lambda\over4} \; g^{ab} \;\nabla_b R.
\end{equation}
(For future convenience, I have chosen a slightly different normalization than Rastall.)
The full Rastall equations of motion (EOM) are then~\cite{Rastall:1972}:
\begin{equation}
G_{ab} + {1\over 4} \lambda\, R \, g_{ab} = \kappa\, [T_\mathrm{\,\sc R}]_{ab},
\end{equation}
whence
\begin{equation}
(\lambda -1 ) \, R = \kappa\, T_\mathrm{\,\sc R}.
\end{equation}
So already at this stage it is clear that the case $\lambda=1$ is special.

There are numerous and extensive claims in the literature that Rastall's approach amounts to introducing a deep non-minimal coupling between gravity and matter.
Unfortunately, as we shall see below, in terms of the underlying physics, this approach proves simply to be a content-free rearrangement of the matter sector. As gravity, there is absolutely nothing new in this proposal.

Similar comments can be found in a little-known 1982 paper by Lindblom and Hiscock~\cite{Lindblom:1982}. As per the discussion below, in this particular 35-year-old article the authors emphasise the construction of a conserved stress-energy tensor, algebraically built from the Rastall stress-energy~\cite{Lindblom:1982}.\footnote{Where Lindblom and Hiscock differ from the current analysis is by introducing the explicit (and quite radical) assumption that laboratory equipment couples only to the non-conserved Rastall stress-energy, not to the conserved stress-energy tensor~\cite{Lindblom:1982}. This allows them to place stringent phenomenological constraints on Rastall's $\lambda$ parameter: $|\lambda| < 10^{-15}$. I will not be exploring this particular route in the current article.
}

\section{Rastall gravity in vacuum }
 
First, we observe that in vacuum Rastall's equation reduces to
\begin{equation}
G_{ab} + {1\over 4} \lambda \,R \,g_{ab} = 0;
\qquad
(\lambda -1 ) \,R = 0.
\end{equation}
If $\lambda\neq 1$ this implies
\begin{equation}
G_{ab}=0;
\end{equation}
whereas if $\lambda=1$ one obtains
\begin{equation}
G_{ab} = \Lambda \, g_{ab}.
\end{equation}
This is either the standard vacuum Einstein EOM,
or at worst the Einstein EOM + (arbitrary cosmological constant).
The vacuum solution is simply an Einstein spacetime.
(For $\lambda\neq 1$ this vacuum degeneracy between the Rastall and Einstein theories was already noted by Rastall some 45 years ago~\cite{Rastall:1972}.)

\section{Adding matter: Generic case ($\lambda\neq 1$)}

Since $R= {\kappa \, T_\mathrm{\,\sc R} \over\lambda-1}$, we construct the geometrical Einstein tensor in terms of Rastall's stress-energy as
\begin{equation}
G_{ab} =  \kappa \left( [T_\mathrm{\,\sc R}]_{ab} + {1\over 4} \;{\lambda\over1-\lambda}\; T_\mathrm{\,\sc R} \; g_{ab}\right).
\end{equation}
Therefore, if we choose to define
\begin{equation}
\label{E:T-decomp}
T_{ab} =  [T_\mathrm{\,\sc R}]_{ab} + {1\over 4} \;{\lambda\over1-\lambda}\; T_\mathrm{\,\sc R} \; g_{ab},
\end{equation}
then this quantity is covariantly conserved.  Thus is is this stress energy that should be considered physical, and in terms of this physical stress-energy tensor 
\begin{equation}
G_{ab} = \kappa \; T_{ab}
\end{equation}
is the usual Einstein equation. 

We can of course invert this construction using
\begin{equation}
T = T_\mathrm{\,\sc R} + {\lambda\over1-\lambda}\; T_\mathrm{\,\sc R} =  {1\over1-\lambda}\; T_\mathrm{\,\sc R}; 
\end{equation}
so that
\begin{equation}
T_\mathrm{\,\sc R} = (1-\lambda) T.
\end{equation}
We see
\begin{equation}
[T_\mathrm{\sc R}]_{ab} =  T_{ab} - {1\over 4} \;{\lambda}\; T \; g_{ab}.
\label{E:T-rastall}
\end{equation}
That is, from the Rastall stress-energy $[T_\mathrm{\,\sc R}]_{ab}$, (and knowledge of the Rastall coupling $\lambda$), one can always reconstruct the physical stress-energy $T_{ab}$,
\emph{and vice versa}.
So, (at least for $\lambda\neq1$), all that is going on is that Rastall has simply mis-identified the physical stress-energy. In terms of the true physical conserved stress-energy $T_{ab}$ one just has standard Einstein gravity.\footnote{Note the existence of an automatic implied consistency condition for Rastall stress-energy: $\nabla_{[c}\nabla^b[T_\mathrm{\,\sc R}]_{a]b} = 0$. 
This might at first glance look ``deep''; unfortunately it is not ``deep''. Observe that one trivially has $\nabla_{[c}\nabla^b[T_\mathrm{\,\sc R}]_{a]b} = {\lambda\over 4} \nabla_{[c} \nabla_{a]} R = 0$.}
%
Indeed, one can easily jump back and forth using equations (\ref{E:T-decomp}) and (\ref{E:T-rastall}).
 Sometimes this very simple observation is hidden very deeply in very technical, very specific, and very turgid calculations.\footnote{For traceless matter, such as electromagnetic stress-energy, the whole process trivializes, $[T_\mathrm{\,\sc R}]_{ab} \to T_{ab}$.}

\section{Adding matter: Special case ($\lambda= 1$)}
This is the only case that is even mildly interesting.
 Ironically, it was already considered (and rejected) by Rastall 45 years ago~\cite{Rastall:1972}.
For $\lambda= 1$ the Rastall EOM reduce to
\begin{equation}
G_{ab} + {1\over 4}  R g_{ab} = \kappa [T_\mathrm{\,\sc R}]_{ab}; \qquad T_\mathrm{\,\sc R} = 0;
\end{equation}
or alternatively
\begin{equation}
R_{ab} - {1\over 4}  R g_{ab} = \kappa [T_\mathrm{\,\sc R}]_{ab}; \qquad T_\mathrm{\,\sc R} = 0.
\end{equation}
So in this $\lambda=1$ special case  situation Rastall matter has to be traceless.
 In terms of the physical stress-energy this is simply
\begin{equation}
G_{ab} + {1\over 4}  R \,g_{ab} = 
\kappa \left(T_{ab} - {1\over 4} T g_{ab}\right),
\end{equation}
or alternatively
\begin{equation}
R_{ab} - {1\over 4}  R \,g_{ab} =
\kappa \left(T_{ab} - {1\over 4} T g_{ab}\right).\quad
\end{equation}
These equations imply that the trace-free part of the Einstein tensor (which equals the trace-free part of the Ricci tensor) is proportional to the trace-free part of the stress-energy tensor. 
This is equivalent to
\begin{equation}
G_{ab}  = \kappa T_{ab} + \Lambda g_{ab}.
\end{equation}
That is, for $\lambda=1$, Rastall gravity  is just ordinary Einstein gravity plus an arbitrary cosmological constant.

Formally this is the same as so-called ``unimodular gravity''~\cite{Ardon:2017,Saez-Gomez:2016,Padilla:2014,Barcelo:2014,Finkelstein:2000,Henneaux:1989}.\footnote{Observe that ``unimodular gravity'' should more properly called ``specified modulus gravity'', meaning that $\det(g)\to\omega$, where $\omega$ is an externally specified and non-dynamical scalar density.} 
Note that for $\lambda=1$ we have\footnote{Even for the special case $\lambda=1$,  there is still an automatic implied consistency condition for the Rastall stress-energy: $\nabla_{[c}\nabla^b[T_\mathrm{\,\sc R}]_{a]b} = 0$. 
This might again at first glance look ``deep''; it isn't. We again trivially have $\nabla_{[c}\nabla^b[T_\mathrm{\,\sc R}]_{a]b} = {1\over 4} \nabla_{[c} \nabla_{a]} R = 0$.}
\begin{equation}
[T_\mathrm{\,\sc R}]_{ab} =  T_{ab} - {1\over 4} T g_{ab};  \qquad T_\mathrm{\,\sc R} = 0.
\end{equation}
So when reconstructing the physical stress-energy one simply has
\begin{equation}
T_{ab} =  [T_\mathrm{\,\sc R}]_{ab} + {1\over 4} T g_{ab};  \qquad T_\mathrm{\,\sc R} = 0.
\end{equation}
That is, from the physical stress-energy $T_{ab}$ you can (uniquely) construct Rastall stress-energy $[T_\mathrm{\,\sc R}]_{ab}$.
In contrast, from the stress-energy Rastall $[T_\mathrm{\,\sc R}]_{ab}$ you can reconstruct the physical stress-energy $T_{ab}$,
up to an \emph{a priori} unknown trace $T$. 
Consequently, even for $\lambda=1$, Rastall gravity is a trivial rearrangement of the
matter sector in Einstein gravity; as gravity there is absolutely nothing new.

\section{Relation of Rastall to trace-free stress-energy}

In terms of the usual stress-energy, let us define the trace-free stress-energy as
\begin{equation}
[T_\mathrm{\,\sc tf}]^{ab}  = T^{ab} - {1\over4}\; T \; g^{ab}. 
\end{equation}
While this trace-free stress-energy tensor certainly shows up in unimodular gravity~\cite{Ardon:2017,Saez-Gomez:2016,Padilla:2014,Barcelo:2014,Finkelstein:2000,Henneaux:1989}, it has a much wider domain of applicability. 

Naturally, this trace-free stress-energy, $[T_\mathrm{\,\sc tf}]^{ab}$, is not (generically) covariantly conserved, indeed we have $\nabla_b [T_\mathrm{\,\sc tf}]^{ab}= -{1\over4} g^{ab} \nabla_b T$, but this covariant non-conservation is not at all a surprise, it is simply due to the way it has been defined. 

Furthermore, since $T^{ab}-[T_\mathrm{\,\sc tf}]^{ab} ={1\over 4} T g^{ab}$, we can always rewrite the Rastall stress-energy of equation (\ref{E:T-rastall}) as a simple linear interpolation between the physical and the trace-free stress-energy tensors:
\begin{equation}
[T_\mathrm{\,\sc R}]^{ab} =  (1-\lambda) T^{ab} + \lambda [T_\mathrm{\,\sc tf}]^{ab}. 
\end{equation}
Non-conservation of Rastall's $[T_\mathrm{\,\sc R}]^{ab}$ is then seen to be an automatic consequence of non-conservation of $[T_\mathrm{\,\sc tf}]^{ab}$; this does not render the Rastall stress-energy any more physical, if anything it further emphasises the purely formal and artificial book-keeping status of the Rastall stress-energy.

\section{Relation of Rastall matter to perfect fluids}

For perfect fluids the inter-relation between Rastall density and pressure, and conserved density and pressure, is simply:
\begin{equation}
\rho_R =   \left(1-{\lambda\over4}\right) \rho + {3\over4}\lambda p; 
\end{equation}
and
\begin{equation}
p_R =\left(1-{\lambda\over4}\right) p -  {1\over4}\lambda\rho.
\end{equation}
Conversely
\begin{equation}
\rho=   \left(1-{\lambda\over4}\over1-\lambda\right) \rho_R - {3\over4(1-\lambda)}\lambda p_R; 
\end{equation}
and
\begin{equation}
p =\left(1-{\lambda\over4}\over1-\lambda\right) p_R -  {1\over4(1-\lambda)}\lambda\rho_R.
\end{equation}
Again we see a simple rearrangement of the matter sector, and no change to the gravity sector. 

\section{Relation of Rastall matter to $w$-matter}

Another quite popular matter model is $w$-matter, where one considers a perfect fluid and defines
\begin{equation}
w = {p\over \rho}; \qquad w_R = {p_R\over \rho_R}.
\end{equation}
However when doing so it is easy to see that
\begin{equation}
w_R = {w+{\lambda\over4}(1-3w)\over1-{\lambda\over4}(1-3w)},
\end{equation}
and conversely
\begin{equation}
\label{E:w(w_R)}
w = {w_R - {\lambda\over4}(1+w_R) \over1-{3\lambda\over4}(1+w_R)}.
\end{equation}
That is, moving back and forward from ordinary $w$-matter to the Rastall version of $w$-matter is simply equivalent to redefining the value of the $w$ parameter and switching $w \longleftrightarrow w_R$. The gravity sector is completely unaffected by this procedure.

Consider for example reference~\cite{Kerr-Rastall:2017}. Those authors consider a Kerr back hole immersed in quintessence ($w$-matter). The only effect of the presence of the $w$-matter is that in the vacuum Kerr spacetime metric one should replace the mass parameter $m$ by 
\begin{equation}
m \to m(r) = m + k\, r^{-3w}.
\end{equation}
(Here $k$ is some constant characterizing the density of the infalling $w$-matter.) If one now considers Rastall-type $w$-matter, then the only \emph{mathematical} change is that one should use equation (\ref{E:w(w_R)}) to simply replace $w\to w(w_R,\lambda)$. 
No actual physics is changing --- the form of the spacetime metric is invariant. One is merely re-parameterizing the physics by using $(w_R,\lambda)$ instead of $w$.

\vspace{-10pt}
\section{Lagrangian/action formulation: lack thereof}

Several authors have noted the absence of any widely accepted and complete Lagrangian formulation (or action formulation) for Rastall  gravity~\cite{MoraisGraca:2017,Ma:2017,Licata:2017,Heydarzade:2017,Bronnikov:2017}, making it at best a phenomenological model, (as we have seen, a trivial phenomenological model). Recently in reference~\cite{Santos:2017} the authors attempted to develop a Lagrangian, but their approach does not fully reproduce Rastall's equations. 

Now one can certainly write a variational principle for the physical conserved stress-energy.
For the usual matter Lagrangian we have
\begin{equation}
\hspace{-20pt}
S_m = \int\sqrt{-g}\, L_{m}\, d^4 x; \quad  T_{ab} = {-2\over \sqrt{-g} } {\delta S_m\over\delta g^{ab}}.\;
\end{equation}
But a similar construction for Rastall stress-energy is lacking. Can one find a ``Rastall action'' $S_\mathrm{\sc R}$ such that 
\begin{equation}
 [T_\mathrm{\,\sc R}]_{ab} = {-2\over \sqrt{-g} } {\delta S_\mathrm{\sc R}\over\delta g^{ab}}\;?
\end{equation}
The key difficulty is this: Since one is attempting to artificially split the total conserved stress-energy into two non-conserved pieces, the price of doing so will be to somehow artificially split the matter action into two pieces that individually lead to non-conserved stress-energies. 
This would require the introduction of  some non-dynamical background field~\cite{Gratus}. 

Even if this could be done, it would not be particularly useful. 
The Einstein--Hilbert action would be unaffected by any such construction --- the gravity sector remains that of standard Einstein gravity. In contrast, the matter sector would be artificially subdivided into two contributions, neither of which is individually covariantly conserved, but whose sum is covariantly conserved. 
We again see that Rastall gravity is simply a repackaging of standard Einstein gravity.
Seeking a Lagrangian/action formulation is not a useful exercise. 

An alternative approach might be to relax what one means by a variational principle, and adopt the modified construction
\begin{equation}
 [T_\mathrm{\,\sc R}]_{ab} = {-2\over \sqrt{-g} } \left\{
 {\delta S_m\over\delta g^{ab}} 
  - {1\over 4} \;{\lambda}\;   g_{ab}  \; g^{cd} \;  {\delta S_m\over\delta g^{cd}} 
\right\}.
\end{equation}
This construction certainly has variational ingredients, but it does not amount to what most people would consider to be a variational principle.

For another example, consider reference~\cite{Corda:2017}.
In that article the authors first split the total action $S$ into an Einstein--Hilbert term $S_{\!EH}$ plus the rest ---  $S_{\!not-EH}$. But then those authors artificially split $S_{\!not-EH}$ into pieces they choose to call $S_{\!matter}$ and  $S_{\!interaction}$. 
Now the standard definition of the stress-energy tensor is based on the metric  variation of  $S_{\!not-EH}$, and immediately leads to a conserved stress-energy tensor. If one artificially splits $S_{\!not-EH}= S_{\!matter}+ S_{\!interaction}$ then neither $T^{ab}_{\!matter}$ nor $T^{ab}_{\!interaction}$ need individually be covariantly conserved --- only their sum need be conserved. But the split into $S_{\!matter}$ and  $S_{\!interaction}$ is only a semantic choice --- no actual physics is involved. Worse, it is not at all clear how their suggested procedure would actually implement Rastall gravity, since they are \emph{ab initio} assuming $T_\mathrm{\,\sc R}^{ab} = T^{ab}_{\!matter}$ then in view of equation (\ref{E:T-decomp}) those authors would need to impose
\begin{equation}
T^{ab}_{\!interaction} =  {1\over 4} \;{\lambda\over1-\lambda}\; T_{matter} \; g^{ab}.
\end{equation}
This is a very strong constraint on their assumed split between the matter and  interaction terms; one that they do not even attempt to justify or discuss in the rest of their article. The rest of reference~\cite{Corda:2017} does not actually address any of the points I have raised in this article; those authors appear to have significantly misinterpreted my actual comments.


\vspace{-10pt}
\section{Discussion}
\vspace{-10pt}

In summary, we have seen that generically Rastall gravity is simply an essentially trivial re-arrangement of the
matter sector in Einstein gravity; as gravity there is absolutely nothing new. Even in the non-generic case, one at best obtains ``unimodular (fixed modulus) gravity'', ordinary Einstein gravity plus an arbitrary cosmological constant. 
It is perhaps sobering to realize that, just because an idea has been in circulation for 45 years, 
does not mean it has been fully debugged. In closing, Rastall gravity is not so much wrong, as it is of rather limited utility.

\vspace{-10pt}
\section*{Acknowledgements}
\vspace{-10pt}
\noindent
This work supported by the Marsden Fund, which is administered by the Royal Society of New Zealand.\\
 I also particularly wish to thank Roberto Percacci, Stefano Liberati,  Ra\'ul Carballo-Rubio, Oliver Piattella, and Hector Okada da Silva for useful comments and discussion. \\
I also wish to thank SISSA (Trieste, Italy) and INFN (Italy) for hospitality.

\vspace{-10pt}

\end{document}